\documentclass[aps,twocolumn,showpacs,floatfix,superscriptaddress,pra]{revtex4}
\usepackage[draft]{hyperref}
\usepackage{amsmath}
\usepackage{amsfonts}
\usepackage{amssymb}
\usepackage{graphicx}

\begin{document}

\title{Semi-discrete solitons in arrayed waveguide structures with Kerr nonlinearity}
\author{N.-C. Panoiu}
\affiliation{Department of Electronic and Electrical Engineering, University College London,
Torrington Place, London WC1E 7JE, UK}
\author{B. A. Malomed}
\affiliation{Department of Physical Electronics, Faculty of Engineering, Tel-Aviv University,
Tel-Aviv 69978, Israel}
\author{R. M. Osgood, Jr.}
\affiliation{Department of Applied Physics and Applied Mathematics, Columbia University, New York,
New York 10027, USA}
\date{\today}

\begin{abstract}
We construct families of optical semi-discrete composite solitons (SDCSs), with one or two
independent propagation constants, supported by a planar slab waveguide, XPM-coupled to a periodic
array of stripes. Both structures feature the cubic nonlinearity and support intrinsic modes with
mutually orthogonal polarizations. We report three species of SDCSs, odd, even, and twisted ones,
the first type being stable. Transverse motion of phase-tilted solitons, with potential
applications to beam steering, is considered too.
\end{abstract}

\maketitle

Power exchange among evanescently coupled cores in arrays of single-mode nonlinear waveguides has
attracted a great deal of interest, because such interactions reveal new phenomena in the wave
dynamics in discrete systems, and due to their potential applications to all-optical switching
(see Ref. \cite{cls03n} for a review of the light transmission in linear and nonlinear discrete
systems). It was predicted that such lattices support discrete solitons
\cite{cj88ol,k93ol,aap96pre}, i.e., localized modes for which the discrete diffraction in the
waveguide array balances the onsite cubic ($\chi ^{(3)}$) nonlinearity, and more complex modes,
such as discrete vortex solitons \cite{mk01pre} (in two-dimensional lattices), discrete surface
solitons \cite{msc05ol} and light bullets \cite{mmf07ol}. In addition, the existence of continuous
solitons in semi-discrete nonlinear media based on Bragg optical fibers have also been recently
predicted \cite{km08josab}. The discrete solitons, as well as their counterparts predicted in
$\chi ^{(2)}$ media \cite{ppl98pre}, have been observed in experiments \cite
{fcs03prl,esm98prl,zwx06prl}.

In recent work \cite{pom06ol}, it was demonstrated that, if a periodic array of optical waveguides
is \emph{nonlinearly} coupled to a slab waveguide, both structures being made of a material with
quadratic ($\chi ^{(2)}$) nonlinearity, the combined optical structure supports a new species of
optical solitons, namely \textit{semi-discrete composite solitons} (SDCS). These solitons are
different from their fully discrete and continuous counterparts in several aspects; in particular,
it has been shown that both even (symmetric) and twisted (antisymmetric) SDCS may be stable.

In this paper, we extend those ideas to the obviously relevant case of composite structures made
of optical media with cubic nonlinearity. We demonstrate, for the first time to our knowledge,
that such media support one- and two-parameter families of stable SDCSs, which consist of two
mutually trapped components, \textit{viz}., a discrete one, which is chiefly carried by the
waveguide array, and a continuous component, bound to the slab waveguide. We also demonstrate that
such SDCS can be used in optically controlled beam steering devices, and therefore they could have
important practical applications.

\begin{figure}[b]
\includegraphics[width=80mm]{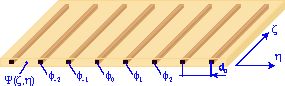} \caption{(Color online) A schematic of the
composite optical structure.} \label{fgeomsol}
\end{figure}

We consider the structure built as a periodic array of stripe optical waveguides, whose adjacent
cores are separated by a distance $d_{0}$, which is coupled to a single-mode slab waveguide. The
waveguide array may be either buried into the slab waveguide, as shown in Fig. \ref{fgeomsol}, or
mounted on top of it. Both the stripes and the slab are to be made of a Kerr material, possible
choices being silicon-on-insulator (SOI) \cite{apl03ol}, polymers \cite{iyi91el}, or semiconductor
heterostructures \cite{tl79jqe}. We also assume that the structure is designed so as to make the
intrinsic modes in the stripe and slab waveguides polarized orthogonally to each other, thus
ruling out the linear coupling between them. Specifically, if we consider buried SOI waveguides,
by simply changing the transverse dimensions of the silicon waveguide one can design stripe
waveguides which support either TE or TM polarized modes \cite{dxs06oe}, and therefore an
experimental set up involving orthogonally polarized modes can be readily implemented. In the
experiment, the incident wide beam may be unpolarized, the stripe and slab waveguides picking
mutually orthogonal polarization components from it. As a result, the propagating mode in each
stripe is linearly coupled to its counterparts in the adjacent ones, and nonlinearly coupled,
through the cross-phase modulation (XPM), to the slab mode. In addition, the case of linear
coupling between a transversely confined waveguide mode and a slab waveguide mode has previously
been considered in Ref. \cite{pa94josab} and therefore it would not be discussed here.

Coupled-mode equations for optical fields in the composite medium can be derived in a consistent
form \cite{js82jqe,m82ke}, resulting in a system of ordinary differential equations for the
discrete component, coupled to a partial differential equation governing the propagation of the
slab mode,
\begin{subequations}
\label{mathModel}
\begin{align}
i\frac{d\phi_{n}}{d\zeta }+& \bar{\beta}_{d}\phi_{n}+\phi_{n-1}+\phi_{n+1}+\phi_{n}|\phi_{n}|^{2}
\notag  \label{psi} \\ & +\kappa_{d}\phi_{n}|\Psi(\zeta ,\eta =n)|^{2}=0, \\
i\frac{\partial \Psi}{\partial \zeta}+& \bar{\beta}_{c}\Psi + \frac{1}{2}\frac{\partial^{2}\Psi
}{\partial \eta^{2}}+\Psi|\Psi |^{2} \notag \\ & +\kappa_{c}\Psi
\sum_{n}|\phi_{n}|^{2}\delta(\eta-n)=0. \label{Psi}
\end{align}
\end{subequations}
Here, $\bar{\beta}_{c,d}=\beta _{c,d}/c_{d}$ are normalized wavenumbers, with $\beta _{c}$
($\beta_{d}$) the propagation constant in the slab (stripe) waveguide in physical units, and
$c_{d}$ the constant of the linear coupling between the stripe modes, while $\zeta =z/z_{0}$ and
$\eta =x/x_{0}$ , where $z$ and $x$ are the longitudinal and transverse coordinates, and
$z_{0}\equiv 1/c_{d}$, $x_{0}\equiv (\beta_{c}c_{d})^{-1/2}$. The normalized fields $\phi_{n}$ and
$\Psi$ are rescaled modal fields $u_{n}$ and $U$ (which are measured in units of
$\sqrt{\mathrm{W}}$ and $\sqrt{\mathrm{W/m}}$, respectively):
$\phi_{n}=\sqrt{\gamma_{d}/c_{d}}u_{n}$ and $\Psi =\sqrt{\gamma_{c}/c_{d}}U$. Finally,
$\kappa_{d}$ and $\kappa_{c}$ are relative strengths of the XPM and SPM\ (self-phase modulation)
interactions, $\kappa_{d}=\sqrt{\gamma_{dc}/\gamma_{c}}$,
$\kappa_{c}=\sqrt{\gamma_{cd}/\gamma_{d}}$. Here, coefficients $\gamma$ are defined as integrals
over cross sections of the respective waveguides ($A_{d}$ and $A_{c}$, the latter defined as the
transverse area of the slab waveguide between adjacent stripes), involving the convolution of the
third-order susceptibility tensor, $\hat{\chi}^{(3)}$, with the stripe and slab modes,
$\mathbb{\mathsf{e}}_{d}(x,y)$ and $ \mathbb{\mathsf{e}}_{c}(y)$, which are normalized to the
corresponding mode powers, $\mathcal{P}_{d}$ and $\mathcal{P}_{c}$ (measured in $\mathrm{W}$ and
$\mathrm{W/m}$):
$\gamma_{d}=(3\epsilon_{0}\omega/4\mathcal{P}_{d}^{2}){\int_{A_{d}}}dA\mathbb{\mathsf{e}}_{d}^{\ast}\cdot
\hat{\chi}^{(3)}\vdots
\mathbb{\mathsf{e}}_{d}\mathbb{\mathsf{e}}_{d}^{\ast}\mathbb{\mathsf{e}}_{d}$,
$\gamma_{c}=(3\epsilon_{0}\omega/4\mathcal{P}_{c}^{2}d_{0}){\int_{A_{c}}}dA\mathbb{\mathsf{e}}_{c}^{\ast}\cdot
\hat{\chi}^{(3)}\vdots\mathbb{\mathsf{e}}_{c}\mathbb{\mathsf{e}}_{c}^{\ast}\mathbb{\mathsf{e}}_{c}$,
$\gamma_{dc}=(3\epsilon_{0}\omega/2\mathcal{P}_{d}\mathcal{P}_{c})
{\int_{A_{d}}}dA\mathbb{\mathsf{e}}_{d}^{\ast}\cdot\hat{\chi}^{(3)}\vdots\mathbb{\mathsf{e}}_{d}\mathbb{\mathsf{e}}_{c}^{\ast}\mathbb{\mathsf{e}}_{c}$,
and $\gamma_{cd}=(3\epsilon_{0}\omega/2\mathcal{P}_{d}\mathcal{P}_{c}d_{0})
{\int_{A_{c}}}dA\mathbb{\mathsf{e}}_{c}^{\ast}\cdot\hat{\chi}^{(3)}\vdots\mathbb{\mathsf{e}}_{c}\mathbb{\mathsf{e}}_{d}^{\ast}\mathbb{\mathsf{e}}_{d}$.
The same system of coupled-mode equations may also be interpreted as the one governing the
transmission of waves $\phi_{n}$ and $\Psi$ with different carrier frequencies, which also rules
out the linear coupling between them.

\begin{figure}[t]
\centerline{\includegraphics[width=8cm]{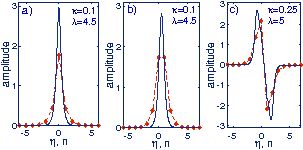}} \caption{(Color online) Field profiles of the
three species of the semi-discrete composite solitons: odd, alias onsite-centered (a); even, alias
intersite-centered (b); and twisted (c).} \label{fsolfoc}
\end{figure}
To focus on the most fundamental case, we assume that the stripe and slab modes are phase-matched,
$\bar{\beta}_{d}=\bar{\beta}_{c}\equiv \bar{\beta}$, hence the corresponding linear terms in Eqs.
(\ref{mathModel}) can by removed (this condition can be easily satisfied by adjusting geometrical
parameters of the stripe and slab), and the nonlinear coupling between the stripe and slab
waveguides is symmetric, with $\kappa_{d}=\kappa_{c}\equiv \kappa$. Then, Eqs. (\ref{mathModel})
conserve the Hamiltonian,
\begin{align*}
H& =-\sum_{n}\left( \phi_{n}^{\ast }\phi_{n+1}+\phi_{n}\phi_{n+1}^{\ast}\right)
-\frac{1}{2}\sum_{n}{\left\vert \phi_{n}\right\vert}^{4} \\ & +\frac{1}{2}\int_{-\infty }^{+\infty
}\left( \left\vert \partial \Psi /\partial \eta \right\vert^{2}-|\Psi|^{4}\right) d\eta -\kappa
\sum_{n}{\left\vert{\phi_{n}}\Psi(n)\right\vert}^{2}
\end{align*}
(the asterisk stands for the complex conjugation), and total powers of the discrete and continuous
components, $P_{d}=\sum_{n}{\left\vert \phi_{n}\right\vert }^{2}$ and
$P_{c}=\int_{-\infty}^{+\infty}{|\Psi|}^{2}d\eta$, respectively.

General soliton solutions to Eqs. (\ref{mathModel}) can be looked for as $ \phi_{n}(\zeta
)=\bar{\phi}_{n}e^{i\lambda_{d}\zeta }$ and
$\Psi(\eta,\zeta)=\bar{\Psi}(\eta)e^{i\lambda_{c}\zeta}$, with $\lambda_{d}\neq \lambda_{c}$,
i.e., they form a two-parameter family of solitons \cite{kmt93pre,hss93pre,hkm03pre}. To find
these solutions, we numerically solved the respective stationary version of Eqs.
(\ref{mathModel}), thus constructing both a particular one-parameter set of the solutions, with
$\lambda_{d}=\lambda_{c}\equiv \lambda$, and the general two-parameter family of SDCSs
($\lambda_{d}\neq \lambda_{c}$).

\begin{figure}[t]
\centerline{\includegraphics[width=8cm]{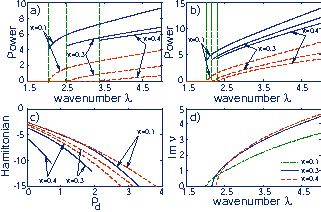}} \caption{(Color online) Top panels: total power
$P$ and power $P_{d}$ of the discrete component (solid and dashed lines, respectively) versus $
\protect\lambda$, for (a) odd and (b) even solitons. (c) Hamiltonian $H$ of the odd and even
solitons (solid and dashed lines, respectively) versus $P_{d}$. (d) The dominant instability
growth rate for even solitons versus $\protect\lambda$.} \label{fpowhaminst}
\end{figure}
Figures \ref{fsolfoc}(a)-(c) display generic examples of the solitons belonging to the
one-parameter families of the \textit{odd}, \textit{even}, and \textit{twisted} types, obtained
through this procedure (the former two may also be classified as \textit{onsite-centered} and
\textit{ intersite-centered}, respectively, as concerns their discrete component). The entire
families are presented in Figs. \ref{fpowhaminst}(a)-(c) by plots showing Hamiltonian $H$, total
power $P\equiv P_{c}+P_{d}$, and the power in the discrete component, $P_{d}$, as functions of
common wavenumber $\lambda$. Both odd and even solitons exist only if $\lambda$ exceeds a
threshold value, $\lambda_{\mathrm{thr}}$, which increases with the coupling strength, $\kappa $.
This result can be explained by the dispersion relation for CW solutions to Eq. (\ref{Psi}), with
amplitude $\Psi_{\mathrm{cw}}$ and transverse wavenumber $k_{0}$, $ \lambda_{\mathrm{cw}}=2\cos
k_{0}+\kappa |\Psi_{\mathrm{cw}}|^{2}$: since nonlinear modes exist for $\lambda >\lambda
_{\mathrm{cw}}$ only, $\lambda_{ \mathrm{thr}}$ indeed increases with $\kappa$. Moreover, diagram
$(H,P_{d})$ in Fig. \ref{fpowhaminst}(c) suggests that the odd (onsite-centered) composite
solitons may be stable, as they correspond to a smaller value of $H$, and become more localized as
$P_{d}$ increases. Indeed, note that the Peierls-Nabarro barrier (PNB), which is defined as the
difference in $H$ of odd and even solitons that have the same power $P_{d}$ \cite{kc93pre},
increases with $P_{d}$.

The expectations concerning the stability of the SDCSs are supported by the linear-stability
analysis. To perform the analysis, we linearized Eqs. (\ref {mathModel}) around the stationary
soliton solutions, $\bar{\phi}_{n}^{ \mathrm{(sol)}}$ and $\bar{\Psi}^{\mathrm{(sol)}}$, setting
$\phi _{n}=\bar{ \phi}_{n}^{\mathrm{(sol)}}+(\delta \phi _{n}e^{i\nu \zeta }+\delta \bar{\phi}
_{n}^{\ast }e^{-i\nu ^{\ast }\zeta })e^{i\lambda \zeta }$ and $\Psi =\bar{
\Psi}^{\mathrm{(sol)}}+(\delta \Psi e^{i\nu \zeta }+\delta \bar{\Psi}^{\ast }e^{-i\nu ^{\ast
}\zeta })e^{i\lambda \zeta }$, and calculating eigenvalues $ \nu $ from the ensuing linearized
equations for the small perturbations. The results, summarized in Fig. \ref{fpowhaminst}(d),
demonstrate that all even solitons are unstable, which is natural to states whose discrete
component is centered at the intersite position \cite{cls03n}. Twisted SDCSs are unstable too, in
their entire domain of existence. On the contrary, odd SDCSs are completely stable, as
$\mathrm{Im}(\nu )=0$ for all $\lambda $ at which they exist.

\begin{figure}[t]
\centerline{\includegraphics[width=80mm]{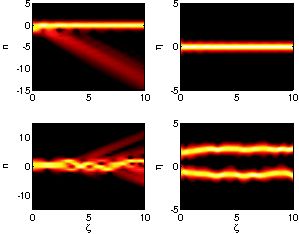}} \caption{(Color online) Unstable propagation
of even (top panels) and twisted (bottom panels) semi-discrete composite solitons. Left and right
panels correspond to the discrete and continuous components, respectively. The soliton parameters
are the same as in Fig. \ref{fsolfoc}.} \label{funstpropfoc}
\end{figure}
A generic scenario for the unstable propagation of even and twisted SDCSs is illustrated in Fig.
\ref{funstpropfoc}, the input solitons being those presented in Fig. \ref{fsolfoc}. Thus, in the
case of even SDCSs, the discrete component evolves into an odd discrete soliton, which then
becomes locked in with the continuous component. As a result, an odd SDCS is formed. Moreover,
during this process part of the energy contained in the initial soliton is released as radiation.
On the other hand, twisted SDCSs break in a pair of mutually trapped odd solitons, which
continuously emit radiation as they copropagate.

\begin{figure}[t]
\centerline{\includegraphics[width=80mm]{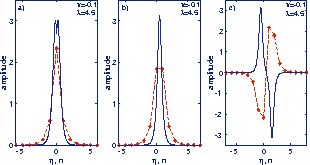}} \caption{(Color online) Field profiles of the
three species of the semi-discrete composite solitons: odd, (a); even, (b); and twisted (c),
corresponding to the case $\kappa<0$.} \label{fsoldfoc}
\end{figure}
\begin{figure}[t]
\centerline{\includegraphics[width=80mm]{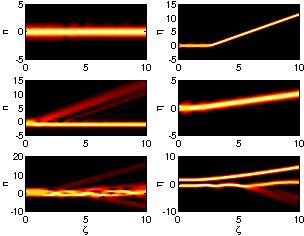}} \caption{(Color online) Unstable propagation
of odd (top panels), even (middle panels), and twisted (bottom panels) semi-discrete composite
solitons. Left and right panels correspond to the discrete and continuous components,
respectively. The soliton parameters are the same as in Fig. \ref{fsoldfoc}.}
\label{funstpropdfoc}
\end{figure}
For the sake of completeness of the study of the model (\ref{mathModel}), we also investigated the
existence and stability of soliton solutions in a more formal case, namely that of $\kappa<0$.
Generic odd, even, and twisted soliton solutions are shown in Fig. \ref{fsoldfoc}. Our analysis
shows that the main difference between this case and the case in which $\kappa>0$ is that for
$\kappa<0$ all three species of solitons are unstable. Thus, as our numerical simulations
illustrate, if $\kappa<0$ odd SDCSs are unstable, too, upon a small transverse perturbation in the
field profile. Specifically, due to the self-defocusing characteristics of the XPM interaction in
this case, the discrete and continuous components repel each other and form discrete and
continuous solitons, which separate from each other upon propagation (see Fig.
\ref{funstpropdfoc}, top panels). This soliton break up is also observed in the case of even and
twisted SDCSs.

\begin{figure}[t]
\centerline{\includegraphics[width=80mm]{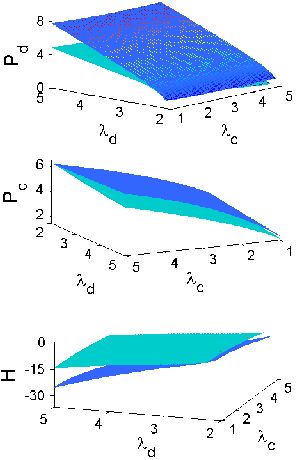}} \caption{(Color online) The power in the
discrete and continuous components (middle and top panels, respectively) of the general soliton
family versus independent wavenumbers $\protect\lambda_{d}$ and $\protect\lambda_{c}$. The
dependence of Hamiltonian $H$ on $\protect\lambda_{d}$ and $\protect \lambda_{c}$ is shown in the
bottom panel. In all panels, the nonlinear-coupling constant is $\protect\kappa=0.1$, the blue and
green (upper and lower) surfaces corresponding to the odd and even SDCS, respectively.}
\label{fphdep}
\end{figure}

General two-parameter SDCS families of the odd and even types, with $\lambda_{d}\neq \lambda_{c}$,
have also been found numerically as stationary localized solutions of Eqs. (\ref{mathModel}). The
results of these calculations, performed for both odd and even SDCSs, are presented in Fig.
\ref{fphdep}, through the respective dependences of the dynamical invariants, $P_{d}$, $P_{c}$ and
$H$, on $\lambda_{d}$ and $\lambda_{c}$. The first among these dependences shows that, similar to
the case of $\lambda_{d}=\lambda_{c}=\lambda$, SDCSs exist only if $\lambda_{d}$ exceeds a certain
threshold value.

To analyze the stability of the general SDCS family, we first used an extension of the
Vakhitov-Kolokolov (VK)\ criterion \cite{vk73rqe} for two-parameter families of solitons
\cite{bkt96prl,bk97prl,pmr03pre}, according to which the stability changes across curves $J=0$ in
the parameter space, where the Jacobian is given by the following expression:
\begin{equation}
J=\frac{\partial (P_{d},P_{c})}{\partial (\lambda_{d},\lambda_{c})}=\frac{\partial P_{d}}{\partial
\lambda _{d}}\frac{\partial P_{c}}{\partial \lambda_{c}}-\frac{\partial P_{d}}{\partial
\lambda_{c}} \frac{\partial P_{c}}{\partial \lambda_{d}}. \label{Jacobian}
\end{equation}
Using the computed powers $P_{d}$ and $P_{c}$, we have concluded that $J\neq 0$ (in fact, $J>0$)
for all values of $\lambda_{d}$ and $\lambda_{c}$ at which SDCSs exist; therefore, both odd and
even SDCSs do not change the VK stability in their existence domains.

Because the VK criterion is only a necessary stability condition [note that it does not predict
the instability of the one-parameter family of even SDCSs, see Fig. \ref{fpowhaminst}(b)], we have
also performed the full linear-stability analysis for the two-parameter families, and concluded
that (as might be expected), the odd solitons are stable, whereas their even counterparts are not.
These conclusions accord to the topology of surface $H(P_{d},P_{c})$, as seen in Fig.
\ref{fphdep}(c): since, for both the odd and even SDCS families, $H(P_{d},P_{c})$ is a smooth
single-valued surface with no cuspidal edges (\textit{Whitney surfaces}), neither type of the
solitons should change its stability within its existence domain \cite{bk97prl,k89pr}.

\begin{figure}[t]
\centerline{\includegraphics[width=80mm]{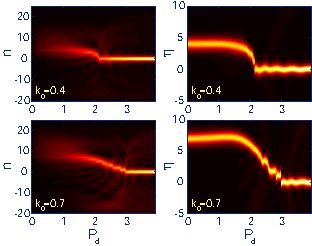}} \caption{(Color online) Contour plots of the
discrete and continuous field components (left and right panels, respectively) of odd (stable)
solitons subjected to the initial phase tilt, $k_{0}$. In this and following figures, the profiles
were recorded after propagation distance $\protect\zeta _{0}=10$ , for coupling constant
$\protect\kappa =0.1$. At $P_{d}<1$, the discrete component is spread over several stripe
waveguides, a transition to pinning being clearly seen at $P_{d}\approx 2$ and $3$, for $
k_{0}=0.4$ and $0.7$, respectively. Note that the continuous component is \emph{stronger
localized} than the discrete one.} \label{fsteering}
\end{figure}
The enhanced localization of the composite solitons with increasing power $P_{d}$ of the discrete
component can be exploited in all-optical beam-steering devices \cite{aap96pre,ppl98pre,pbo03ol}.
To demonstrate this possibility, we consider SDCSs with a phase tilt, accounted for by transverse
wavenumber $k_{0}$, at the input facet of the waveguide array, $\left\{\phi_{n},\Psi \right\}
|_{\zeta =0}=\left\{ \bar{\phi}_{n}^{\mathrm{(sol)}},\bar{\Psi}^{\mathrm{(sol)}}\right\}
e^{ik_{0}\eta }$, and follow their propagation in the waveguide array. Figure \ref{fsteering}
presents the results for odd (stable) solitons, in the case of $\lambda_{d}=\lambda_{c}=\lambda$.
As might be expected, at low power $P_{d}$ the \textquotedblleft kicked" (phase-tilted) SDCS
readily moves across the waveguide array. However, as $P_{d}$ increases, the transverse
displacements of both the discrete and continuous components decrease, and, for $P_{d}$ exceeding
a certain critical value [in Fig. \ref{fsteering}, $\left( P_{d}\right) _{\mathrm{cr}}\approx 2$
and $3$, for $k_{0}=0.4$ and $0.7$, respectively], the soliton remains trapped at its initial
location. This finding is consistent with the fact that the PNB increases with $P_{d}$, as per
Fig. \ref{fpowhaminst}(c).

\begin{figure}[t]
\centerline{\includegraphics[width=80mm]{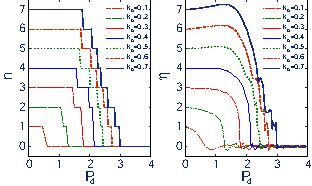}} \caption{(Color online) The location of the
peak intensity in the discrete and continuous components (left and right panels, respectively) of
the odd soliton versus power $P_{d}$ of its discrete component, at different values of phase pitch
$k_{0}$ applied at $\protect\zeta =0$. The coupling constant is $\protect\kappa =0.1$.}
\label{fkdep}
\end{figure}

Dependencies of the transverse displacement of the odd SDCS on phase pitch $k_{0}$, which causes
its displacement, and power $P_{d}$ of its discrete component, are displayed in Fig. \ref{fkdep}.
Naturally, the transverse displacement increases with $k_{0}$, as well as critical power $\left(
P_{d}\right)_{\mathrm{cr}}$ required to pin the soliton. It is worthy to note that the discrete
and continuous components remain mutually trapped even close to the linear limit (at small
$P_{d}$), which is not obvious in Fig. \ref{fsteering}.

To summarize, we have demonstrated that the structure built as a single-mode slab waveguide
coupled to a discrete waveguide array, with both parts made of a Kerr-nonlinear material, supports
semi-discrete composite solitons of odd (onsite-centered), even (intersite-centered), and twisted
types. One- and two-parameter families of the odd solitons (with one or two independent
wavenumbers, respectively) are stable, while the other types are not. Note that in the case of
optical media with quadratic nonlinearity both even and twisted SDCSs may be stable. This should
not be a surprise as quadratically and cubically nonlinear media represent essentially different
optical system and therefore one expects that they support nonlinear modes with different physical
properties. In addition, in the case of quadratically nonlinear media, SDCSs have a more complex
structure, as the discrete and the continuous components can be excited at either the fundamental
frequency or at the second harmonic.

The potential application of the stable solitons, kicked in the transverse direction, to
all-optical beam steering was also demonstrated. The analysis presented here for the
one-dimensional setting can be extended to two-dimensional lattices and photonic-crystal fibers.
For example, nonlinear modes supported by 2D nonlinear optical media invariant to discrete
symmetry transformations have already been investigated \cite{mks00pre,pbo04oe,smb05oe,skp05sap}.
It may also be interesting to analyze semi-discrete solitons in media with the self-defocusing
cubic nonlinearity, where the discrete component would be of the \textit{staggered} type.

This work was supported by NSF under Grant No. ECS-0523386, and by DARPA/AFOSR under Grant No.
FA9550-05-1-0428.


\begin{thebibliography}{99}
\bibitem{cls03n}D. N. Christodoulides, F. Lederer, and Y. Silberberg, \nat \textbf{424}, 817
(2003).

\bibitem{cj88ol}D. N. Christodoulides and R. I. Joseph, \ol \textbf{13}, 794 (1988).

\bibitem{k93ol}Y. S. Kivshar, \ol \textbf{18}, 1147 (1993).

\bibitem{aap96pre}A. B. Aceves, C. De Angelis, T. Peschel, R. Muschall, F. Lederer, S. Trillo, and
S. Wabnitz, \pre \textbf{53}, 1172 (1996).

\bibitem{mk01pre}B. A. Malomed and P. G. Kevrekidis, \pre \textbf{64}, 026601 (2001).

\bibitem{msc05ol}K. G. Makris, S. Suntsov, D. N. Christodoulides, G. I. Stegeman, and A. Hache,
\ol \textbf{30}, 2466 (2005).

\bibitem{mmf07ol}D. Mihalache, D. Mazilu, F. Lederer, and Y. S. Kivshar, \ol \textbf{32}, 2091
(2007).

\bibitem{km08josab}K. Levy and B. A. Malomed, \josab \textbf{25}, 302 (2008).

\bibitem{ppl98pre}T. Peschel, U. Peschel, and F. Lederer, \pre \textbf{57}, 1127 (1998).

\bibitem{fcs03prl}J. W. Fleischer, T. Carmon, M. Segev, N. K. Efremidis, and D. N.
Christodoulides, \prl \textbf{90}, 023902 (2003).

\bibitem{esm98prl}H. S. Eisenberg, Y. Silberberg, R. Morandotti, A. R. Boyd, and J. S. Aitchison,
\prl \textbf{81}, 3383 (1998).

\bibitem{zwx06prl}H. Zeng, J. Wu, H. Xu, and K. Wu, \prl \textbf{96}, 083902 (2006).

\bibitem{pom06ol}N. C. Panoiu, R. M. Osgood, and B. A. Malomed, \ol \textbf{31}, 1097 (2006).

\bibitem{apl03ol}V. R. Almeida, R. R. Panepucci, and M. Lipson, \ol \textbf{28}, 1302 (2003).

\bibitem{iyi91el}S. Imamura, R. Yoshimura, and T. Izawa, Electron. Lett. \textbf{27}, 1342 (1991).

\bibitem{tl79jqe}W. T. Tsang and R. A. Logan, \jqe \textbf{15}, 451 (1979).

\bibitem{dxs06oe}E. Dulkeith, F. Xia, L. Schares, W. M. J. Green, and Y. A. Vlasov, Opt. Express
\textbf{14}, 3853 (2006).

\bibitem{pa94josab} K. P. Panajotov and A. T. Andreev, \josab \textbf{11}, 826 (1994).

\bibitem{js82jqe} S. M. Jensen, \jqe \textbf{18}, 1580 (1982); G. I. Stegeman, \jqe \textbf{18},
1610 (1982).

\bibitem{m82ke}A. M. Maier, Kvantovaya Elektron. (Moscow) \textbf{9}, 2296 (1982); [Sov. J. Quant.
Electr. \textbf{12}, 1490 (1982)].

\bibitem{kmt93pre}D. J. Kaup, B. A. Malomed, and R. S. Tasgal, \pre \textbf{48}, 3049 (1993).

\bibitem{hss93pre}M. Haelterman, A. P. Sheppard, and A. W. Snyder, \ol \textbf{18}, 1406 (1993).

\bibitem{hkm03pre}J. Hudock, P. G. Kevrekidis, B. A. Malomed, and D. N. Christodoulides, \pre
\textbf{67}, 056618 (2003).

\bibitem{kc93pre}Y. A. Kivshar and D. K. Campbell, \pre 48, 3077 (1993).

\bibitem{vk73rqe}M. G. Vakhitov and A. A. Kolokolov, Radiophys. Quant. Electr. \textbf{16}, 783
(1973).

\bibitem{bkt96prl}A. V. Buryak, Y. S. Kivshar, and S. Trillo, \prl \textbf{77}, 5210 (1996).

\bibitem{bk97prl}A. V. Buryak and Y. S. Kivshar, \prl \textbf{78}, 3286 (1997).

\bibitem{pmr03pre}N. C. Panoiu, D. Mihalache, H. Rao, and R. M. Osgood, \pre \textbf{68},
065603(R) (2003).

\bibitem{k89pr}F. V. Kusmartsev, Phys. Rep. \textbf{183}, 1 (1989).

\bibitem{pbo03ol}N. C. Panoiu, M. Bahl, and R. M. Osgood, \ol \textbf{28}, 2503 (2003);
\textit{ibid.}, \josab \textbf{21}, 1500 (2004).

\bibitem{mks00pre}S. F. Mingaleev, Y. S. Kivshar, and R. A. Sammut, \pre \textbf{62}, 5777 (2000).

\bibitem{pbo04oe}N. C. Panoiu, M. Bahl, and R. M. Osgood, Opt. Express \textbf{12}, 1605 (2004).

\bibitem{smb05oe}G. Van der Sande, B. Maes, P. Bienstman, J. Danckaert, R. Baets, and I.
Veretennicoff, Opt. Express \textbf{13}, 1544 (2005).

\bibitem{skp05sap}J. R. Salgueiro, Y. S. Kivshar, D. E. Pelinovsky, V. Simon, and H. Michinel, St.
Appl. Math. \textbf{115}, 157 (2005).
\end{thebibliography}
\end{document}